\documentclass{SCIS2020}

\begin{document}
\ArticleType{REVIEW}
\Year{2020}
\Month{}
\Vol{}
\No{}
\DOI{}
\ArtNo{}
\ReceiveDate{}
\ReviseDate{}
\AcceptDate{}
\OnlineDate{}

\title{A Survey of Prototype and Experiment for UAV Communications}{A Survey of Prototype and Experiment for UAV Communications}

\author[1,2]{Qingheng Song}{}
\author[3]{Yong Zeng}{{yong\_zeng@seu.edu.cn}}
\author[4]{Jie Xu}{}
\author[3]{Shi Jin}{}

\AuthorMark{Qingheng Song}

\AuthorCitation{Q. Song, Y. Zeng, J. Xu and S. Jin}
\address[1]{College of Electrical and Information Engineeing, Huaihua University, Huaihua {\rm 418008}, China}
\address[2]{Key Laboratory of Intelligent Control Technology for Wuling-Mountain Ecological Agriculture in Hunan\\ Province, Huaihua {\rm 418008}, China}
\address[3]{National Mobile Communications Research Laboratory, Southeast University, Nanjing {\rm 210096}, China}
\address[4]{School of Science and Engineering, The Chinese University of Hong Kong, Shenzhen  {\rm 518172}, China}

\abstract{Unmanned aerial vehicle (UAV) communications have attracted significant attention from both academia and industry. To facilitate the large-scale usage of UAVs for various applications in practice, we provide a comprehensive survey on the prototype and experiment for UAV communications. To this end, we first provide an overview on the general architecture of the prototype and experiment for UAV communications, and then present experimental verification for air-to-ground channel models and UAV energy consumption models. Next, we discuss measurement experiments on two promising paradigms of UAV communications, namely cellular-connected UAVs and UAV-enabled aerial communication platforms. For the former, we focus on the feasibility study and address the interference mitigation issue. For UAV-enabled aerial communication platforms, we present three scenarios, namely UAV-enabled aerial base stations, UAV-enabled aerial relays and UAV-enabled aerial data collection/dissemination. Finally, we point out some promising future directions for prototype and experimental measurements for UAV communications.} 

\maketitle

\section{Introduction}
With technological advancements, unmanned aerial vehicles (UAVs) have recently found a wide range of civilian and commercial applications, such as aerial surveillance, cargo transportation, search and rescue, pollution monitoring, agriculture, film-making, and wireless communications and networks. In particular, thanks to the three dimensional (3D) mobility, on-demand and swift deployment capability, as well as the possession of line-of-sight (LoS) air-to-ground (A2G) communication links, UAVs are expected to play an important role in future wireless networks, such as Internet of things (IoT), wireless sensor networks (WSN), poster-disaster communication recovery, and data offloading of hotspots \cite{ZengWireless, ZengAccessing}.
In general, the typical applications of UAV communications can be classified into UAV-enabled communication platforms and cellular-connected UAV communications based on service providing manner. Furthermore, UAVs can be loosely classified into two categories based on wing type, namely fixed- and rotary-wing UAVs, respectively. Rotary-wing UAVs can hover at fixed locations in the air and flexibly change their flight directions, but normally with quite limited payload capability \cite{JohnsonHelicopter}. In contrast, fixed UAVs must maintain a forward flight to remain aloft, but usually can carry more payload and are more energy-efficient due to their gliding characteristic \cite{FilipponeFlight}.

To facilitate the design of UAV communication systems, various methods such as theoretical analysis, computer-based simulations, and prototyping or experimental measurements can be performed to evaluate or validate the resulting performance, each of which has its own advantages and disadvantages. For example, theoretical analysis can generally offer useful insights and guidelines to the design and performance optimization of the UAV communication systems, but they usually rely on certain idealized assumptions for analytical tractability, thus compromising the feasibility and achievability in practice. On the other hand, computer-based simulations can be used to imitate the real environment (e.g., wireless channels, building shapes and heights, user distributions, and user behaviors) for UAV communications, to help obtain general design insights. Although the computer-based simulations are usually of low cost, the practicality of their results highly depends on the modelling accuracy. By contrast, the prototyping and experimental measurements can avoid the modelling bias associated with simulation and theoretical methods, identify the technical issues and challenges that are overlooked in theoretical analysis and simulations, and bridge the gap between theory and practice to help accelerate the commercialization of various  UAV communications. However, experimental measurements or prototypes are generally of high cost and time consuming in practical implementation. Therefore, the aforementioned three methods are usually complementary with each other.

Note that in the literature, there are extensive works on the survey and overview of UAV communications from different aspects (see, e.g., \cite{ZengWireless, ZengAccessing, WalidBOOK, Hayatcivil, Shakhatrehcivil,  MozaffariATutorial, ChengMEC, GuptaUAVnetworks, MotlaghIOT, Jiangrout, LuWPT, CaoAirborne, ZhangUAV-to-X, MozaffariD2D, BerghLTE, ZengCellular, FotouhiCellular, BekmezciFAHOC, LinCellular, Cachingchen}).  For instance, the authors in \cite{ZengWireless} provided an overview on UAV-enabled communication platforms, by presenting three typical use cases including ubiquitous coverage enhancement, mobile relaying and information collection/dissemination, together with the corresponding opportunities and challenges.
The authors in \cite{ZengAccessing} presented a tutorial on UAV communication for 5G-and-beyond wireless networks, in which the state-of-the-art results for UAV-enabled communication platforms and cellular-connected UAV communications are reviewed, and the fundamentals on performance analysis, evaluation, and optimization for UAV communications are also presented.
Furthermore, the book \cite{WalidBOOK} discussed various issues on UAV communications such as performance analysis and optimization, physical layer design, trajectory and path planning, resource management, multiple access, cooperative communications, standardization, control, and security;
and \cite{MozaffariATutorial} presented a comprehensive tutorial on the applications of UAVs in both aerial base station and cellular-connected UAVs, in which the important challenges, the fundamental tradeoffs, and open problems are discussed. Besides, there are some other works reviewing specific UAV applications in civil applications \cite{Hayatcivil, Shakhatrehcivil}, ad-hoc networks \cite{GuptaUAVnetworks, Jiangrout, CaoAirborne, BekmezciFAHOC}, IoT \cite{MotlaghIOT}, D2D communications \cite{MozaffariD2D}, UAV-to-X \cite{ZhangUAV-to-X}, mobile edge networks \cite{ChengMEC}, caching \cite{Cachingchen}, wireless power transfer (WPT) \cite{LuWPT},
and cellular-connected UAV  users \cite{BerghLTE, ZengCellular, LinCellular, FotouhiCellular}. However, these prior works mainly focused on the theoretical analysis and computer-based simulations, while to our best knowledge, a comprehensive overview on UAV communications from the prototype and experiment perspectives is still lacking. This thus motivates our current work to fill such a gap. 

To develop prototype or experimental measurements for UAV communications, one first needs to construct a UAV platform, by selecting the proper aircrafts, communication technologies and network protocols. The experimental platforms can be constructed via adopting Commercial-Off-The-Shelf (COTS) solutions, customized solutions or a combination of the two. Furthermore, to evaluate the performance of UAV communications, the A2G channel characteristics and the limited size, weight and power (SWAP) issues of UAVs are of paramount importance. Therefore, it is important to properly model the A2G wireless channels and the UAV energy consumption to lay the foundation to design UAV communication systems.
Furthermore, thanks to the device miniaturization of communication equipment, the continuous cost reduction in UAV manufacturing, the fast development of durable and light weight manufacture material, and the advancement of computing, communication, and sensing units, it becomes feasible to mount compact communication equipment on UAVs to enable the integration of UAVs into terrestrial communication networks. There are two typical paradigms for such an integration, namely, cellular-connected UAV, where UAVs with their own missions are connected to terrestrial networks as aerial users,  and UAV-enabled aerial platform, where dedicated UAVs are deployed as aerial base stations, relays, or access points to assist the terrestrial communications from the sky.

In this paper, we present a comprehensive survey on the prototypes and experiments for UAV communications. The main contributions of this paper are summarized as follows.
\begin{itemize}
\item First, we provide an overview on the general architecture for UAV communication prototypes and experiments, including aircraft selection, communication technologies, and communication protocols.
\item Next, we present an extensive overview on experiments for UAV A2G channel modeling and energy consumption modeling.
\item Then, we review the existing measurement campaigns on cellular-connected UAVs, investigate the feasibility of cellular-connected UAVs, and discuss the promising technologies to mitigate the A2G interference.
\item Furthermore, we present an overview on experiments for UAV-enabled communication platforms, including three typical use cases, i.e, UAV-enabled aerial base station, UAV-enabled aerial relay and UAV-enabled data collection/dissemination.
\item Finally, we point out some important scenarios of prototype and experiment for UAV communications that deserve further investigation in future work.
\end{itemize}

The rest of this paper is organized as follows. The general architecture for UAV prototype and experiment, experiments for A2G channel modeling and UAV energy consumption modelling are presented in section 2, 3, and 4, respectively. Section 5 and Section 6 discuss the prototypes and experiments for cellular-connected UAVs and UAV-enabled aerial communications platforms, respectively. The future trends of prototype and experimental measurements for UAV communications are presented in Section 7. Finally, Section 8 concludes the paper.

\section{General architecture for UAV communication prototype and experiment}
\begin{figure}[H]
\centering
\includegraphics[width=5.5cm]{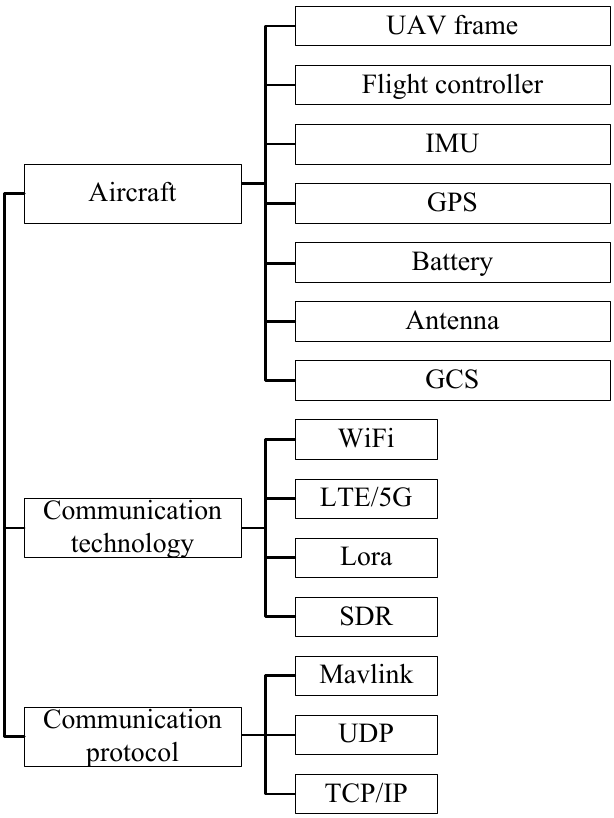}
\caption{Major components for UAV experiment.}
\label{UAV_Structure}
\end{figure}

\begin{figure}[tp]
\centering
\includegraphics[width=10cm]{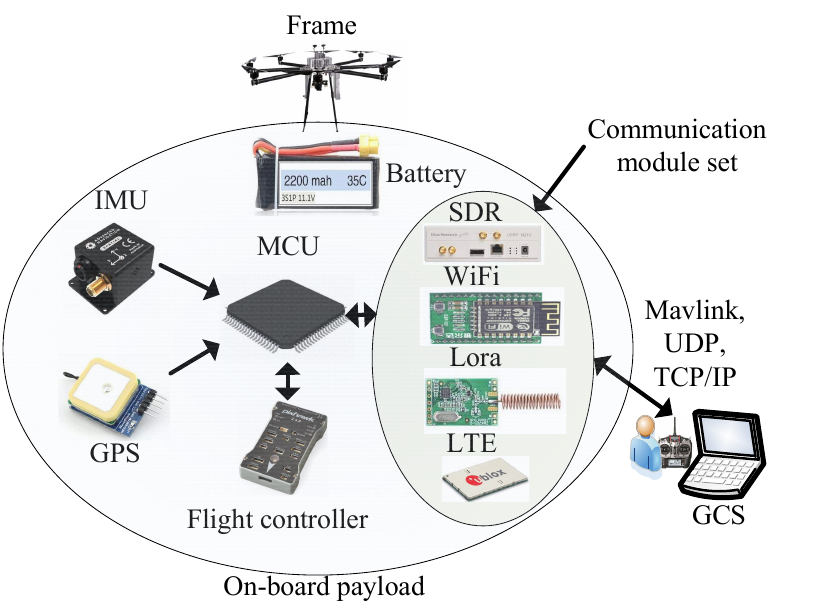}
\caption{Illustration of typical prototype construction for UAV experiment.}
\label{aircraft}
\end{figure}

For the general architecture of standalone (for most current UAVs) and networked UAV communication (for future UAVs) experiment, we first discuss the aircraft selection, and then, we provide an overview of communication technologies, including WiFi, LoRa, long-term evolution (LTE), and customized soft-defined radio (SDR)-based technology. Finally, the UAV communication protocols are presented. The major components  for UAV experiment is summarized in Figure \ref{UAV_Structure}, and an illustration of a typical prototype construction for UAV experiment is given in Figure \ref{aircraft}.
\subsection{Aircraft selection}
The practical performance of UAV communications can be better understood via field experiments. Researchers may construct their own prototypes by choosing and developing suitable hardware and software to implement a specific mission. The hardware includes the vehicle, flight avionics (e.g., the flight controller, Global Positioning System (GPS), and inertial measurement unit (IMU)), antenna, battery and communications-related equipment. The flight controller is responsible for the stable flight of UAV, usually with predefined waypoint navigation. The GPS is used for localization, and the IMU (e.g., compass, gyroscope, magnetometer and accelerometer) is utilized to measure flight dynamics such as pitch, yaw and roll angles. The communication-related equipment is used to establish communication links between UAV and ground control stations (GCSs), ground users, or terrestrial base stations, which include both command \& control and payload communication links. In addition, the software is often developed for implementation of communications-related functions, and GCS functions, such as mission planning, monitoring, data exchanging, and control.
System level design of an UAV communication system requires an all-COTS solution or a completely customized design, or a combination of both. To construct a vehicle, several major factors should be considered, e.g., payload capability, endurance, expandability. From a practical perspective, different applications usually require different types of UAVs due to specific requirements in terms of payload, endurance, operating environment, and cost, etc. Compared with terrestrial communication systems which are usually powered by grid or uninterruptible power systems, the SWAP constraints of UAVs raise serious limits on their payload and endurance capabilities. UAVs may not be able to carry large-capacity rechargeable battery, and heavy, bulky, and energy-consuming communication-related equipments, such as high-performance signal processing equipment, and dish antenna. The UAV-enabled communications systems, e.g., base stations or relays, generally require lighter weight and more compact hardware design compared to terrestrial communications systems. As a result, it needs to be carefully designed to cater for the SWAP constraint of UAVs. For example, the airframes of fixed-wing UAVs are usually constructed by using foam materials, and that of rotary-wing UAVs are constructed by using durable and lightweight carbon fibre or aluminum alloy.

In addition, both propulsion power and communication-related power of UAVs are provided by the onboard battery. First, the endurance of UAVs highly depends on the total weight of UAV system and capacity of onboard battery. According to \cite{DochevEmbedding}, increasing battery capacity results in a proportional increase in battery weight, but the endurance may not increase proportionally since part of the increased power is dedicated for the increased weight.
Therefore, the selection of onboard battery should consider weight, power, and cost.
The battery technology has been continuously advanced to enable more onboard energy storage on the same battery weight, and several advanced techniques have also been actively studied, such as solar-powered UAVs \cite{WilkinsCritical}, wireless power transfer \cite{LuCharging}, and laser-powered UAVs \cite{LuCharging, OuyangThroughput, Zhaolaser}. For the expandability, additional space, signal processing ability, payload capability remained and interfaces are needed to install various sensors or equipments for future expansion.

Several open source projects can also be directly used to build both fixed- and rotary-wing UAV from a customized manner, such as Paparazzi \cite{paparazzi}, Ardupilot \cite{ardupilot}, Openpilot \cite{openpilot}, and Pixhawk \cite{pixhawk}. One of the main reasons to use open source projects is their flexibility in both hardware and software, including UAV frame, flight avionics, antenna design, and GCS, which makes their modification and development easier and quicker to meet the specific requirements. In addition, open source projects allow developers to directly utilize and extend the results of others.
An overview of available open source projects for UAV is presented in \cite{LimBuild}, where eight open source projects were presented with descriptions about their avionics, sensors, attitude estimation, and control algorithms. By using these open source projects, researchers can develop their own UAV systems at a relatively low cost with less efforts.

The aforementioned methods are mainly for the customized designs. The COTS solutions can also be adopted by UAV developers worldwide, such as DJI Wookong, DJI Matrice 100, Parrot ARDrone, and Draganflyer X4, which can significantly shorten the development period. The researchers can directly use these COTS platforms to carry some communication equipments or develop specified communication functionalities using software development kit (SDK) provided by the manufacturers. The COTS solutions are ready to fly once open box, including aircraft frame, motor (motor driver), propeller, flight controller, GPS, battery, GCS, equipment for command \& control information and data exchanging, etc. The COTS solutions also have some other advantages. For example, DJI products are programmable by using the DJI SDK, which supports Linux, Ros, QT, and embedded systems. The DJI products can be controlled remotely via DJI GO installed in a laptop. Some critical information can be displayed, e.g., real-time trajectory, battery time, and GPS signal strength. In addition, the DJI products provide universal power and communications ports, including CAN ports and UART ports, which allow other communication modules to be connected with DJI products. However, the DJI products solutions may not satisfy the specific requirement with various functions. One solution is to develop customized functions by using DJI SDK, which, however, is time-consuming and rather demanding on developers' skills. Another solution is to install additional communication equipments on the UAV, such as LTE equipment, WIFI, and LoRa, which will increase the payload of UAV, and compromise endurance significantly. It is worth mentioning that DJI products mainly focus on rotary-wing UAV, and the development for fixed-wing UAV is still ongoing. Another disadvantage of COTS solutions is their high cost which constrains their applications severely. In particular, we provide a comparison of different COTS UAVs in Table \ref{Aircraftselection}, which can be directly adopted for prototype and experiment on UAV communications.
\begin{table}[H] \footnotesize
\centering
\renewcommand{\arraystretch}{1.5}
\caption{Comparison of different COTS UAVs}
\label{Aircraftselection}
\begin{tabular}{|p{1.3cm}|p{0.8 cm}|p{1.1cm}|p{1.6cm}|p{1 cm}|p{1.5 cm}|p{2 cm}|p{1.4 cm}|p{1.4 cm}|}
  \hline
  \textbf{UAV} &\textbf{Wing} &\textbf{Manuf-acturer}&\textbf{Take off condition} & \textbf{Mater-ial} & \textbf{Endurance} & \textbf{Weight/ Payload}& \textbf{Maximum speed} & \textbf{Frequency band}\\
  \hline
  DJI Wu Inspire 2 &  Rotary & DJI & Unrestricted & Alloy, carbon fiber & 27 minutes& 3.44kg/0.81kg& 26 m/s& 2.4GHz, 5.8GHz\\
  \hline
  DJI Matrice 200 &  Rotary & DJI & Unrestricted & Alloy, carbon fiber & 38 minutes& 4.69 kg/1.45 kg& 22.5 m/s& 2.4GHz, 5.8GHz\\
  \hline
  P550H &  Rotary & CHCNAV & Unrestricted & Carbon fiber & 64 minutes& 6.2 kg/8 kg & -& Unspecified\\
  \hline
  Cumulus One &  Fixed & Sky-Watch & Hand launched & Foam & 2.5 hours & -/0.5kg & 16.1 m/s& Unspecified\\
  \hline
  UX11 &   Fixed & Delair & Hand launched & Foam & 59 minutes& 1.5kg(including payload) & 15 m/s& Unspecified\\
  \hline
  DATAhawk &  Fixed & Quest-UAV & Hand launched & Carbon Fibre & - & 2.15kg(including payload) & 27.78 m/s& 2.4 GHz, 868 MHz\\
  \hline
\end{tabular}
\end{table}

Finally, researchers may also build their aircraft by utilizing a combination of customized and COTS solution. For example, one can use the frame of COTS solution by upgrading the motor and driver using customized solution to enhance the payload capability. In addition, since some sensors may not be provided by the COTS solution (e.g., wind speed sensors), the users can install additional sensors and corresponding circuits on the COTS aircraft. The combination of customized and COTS solution provides a flexible tradeoff between development period and cost.

\subsection{Communication technology}
The widely used communication technologies for both command \& control and payload communications include WiFi, LoRa, LTE, and customized SDR-based communication technology. 

For commercial UAVs today, WiFi is the most popular communication technology \cite{NunnsAutonomous, VilharExperimental, GuAirborne}, which usually operates at unlicensed band (2.4 GHz, 5 GHz) and is supported by many commercial companies and open source projects, due to the cost, regulation, and compatibility considerations. The technology is based on the IEEE 802.11 standards and utilizes the carrier sense multiple access (CSMA) collision avoidance mechanism. WiFi equipment can be directly obtained off-the-shelf without the need for new design and modification.
In \cite{NunnsAutonomous}, a design and implementation of flying WiFi access point was presented by using commercial components including Raspberry Pi, Crius CN-06 GPS module, DJI 2312E motors, One Pro Flight Controller, and DJI F450 frame. The tests were performed at a outdoor park. The Raspberry Pi was set to a static WiFi channel. Their evaluation demonstrated that the UAV can adjust their position adaptively to provide best signal strength for ground users. %
In \cite{VilharExperimental}, two bands were both tested in the experiment with a balloon and a car by equipping with WiFi, where the access link was established at 2.4 GHz and the backhaul link was established at 5 GHz. The GPS data as well as Received Signal Strength Indicator (RSSI) were recorded. The results showed that the range of WiFi is limited by the transmit power and the receiver sensitivity.
In \cite{GuAirborne}, an experimental study of airborne WiFi networks with directional antenna was implemented to study the feasibility of transmitting WiFi signals over two UAVs. The design included payload adjustment, directional antenna and a mechanical heading control. The DJI F550 hexacopter and NAZA-M Lite flight controller were adopted with a 7500 mAh LiPO battery installed on the UAV. The gyroscope, accelerometer, GPS and barometer were integrated in NAZA-M system. The effect of distance on the throughput was measured.

LoRa which operates at 868 MHz or 915 MHz is another promising communication technology for UAV communications due to its features of low power and long-range \cite{DambalImproving, GodoyLoRa}. It can achieve an adjustable data rate by varying the spreading factor. The theoretical coverage area of LoRa is 15 km for suburban area and 5 km for urban area. In \cite{GodoyLoRa}, the performance of LoRa-enabled UAV communications was measured, e.g., communication distance, transmit power, signal-to-noise ratio and the RSSI. The LoRa transceiver used is the Semtech SX1272 module designed by Libelium. The experiments validated that LoRa can reach a communication distance of 10 km using a transmit power of 0 dBm. In \cite{DambalImproving}, the signal strength measurements were performed in urban and suburban environments by deploying the LoRa transmitter on a DJI Phantom 4 Pro at different altitude between 25 m and 50 m. They found that the altitude and antenna orientation were critical for coverage.

In addition, cellular-connected UAVs through LTE technologies have many promising advantages, such as almost ubiquitous cellular infrastructures, which can significantly extend the communication range between UAV and GCS to beyong visual line-of-sight link. Cellular technologies can offer the connectivity for UAVs as proposed in the 3rd generation partnership project (3GPP) work item for UAVs operating on LTE \cite{Muruganathan3GPP}. In \cite{Muruganathan3GPP}, the authors investigated the capability of LTE networks for providing connectivity to UAVs. They provided an overview of the key findings of the 3GPP Release-15, introduced the connectivity requirements and performance evaluation scenarios, discussed the channel models and challenges.

\begin{figure}[H]
\centering
\includegraphics[width=6cm]{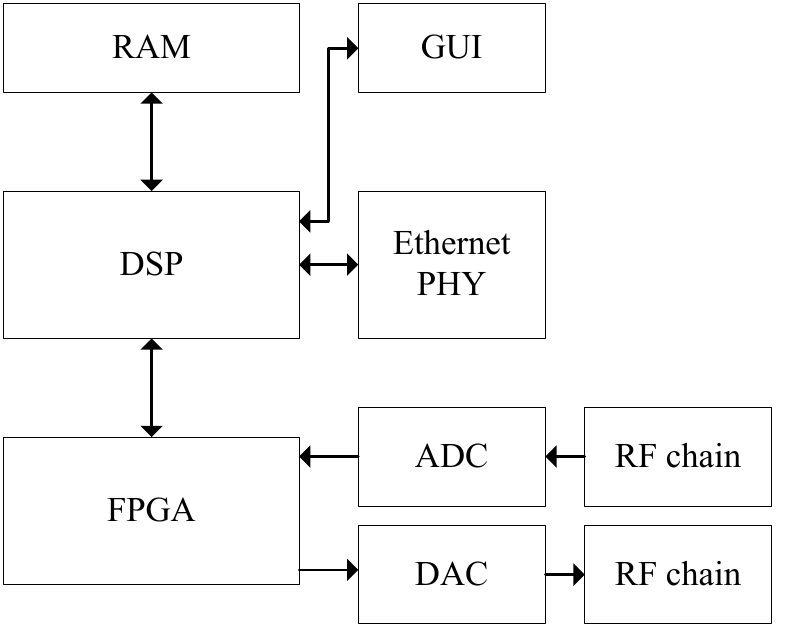}
\caption{Components of a typical SDR.}
\label{SDR}
\end{figure}

SDR is a flexible and low-power radio technology that can be used to develop the UAV communications platform. The authors in \cite{PowellSDR} provided a comprehensive overview of SDR from both hardware and software perspectives, which can be used  for UAV wireless experimentation and research.
The authors in \cite{MarojevicPlatform} introduced a SDR-based aerial experimentation and research platform for advanced wireless, and presented an architecture for designing prototype to enable controllable aerial experiments with latest wireless technologies and systems.
The diagram of a typical SDR is given in Figure \ref{SDR}, which generally consists of digital signal processor (DSP), field programmable gate array (FPGA), random-access memory (RAM), graphical user interface (GUI), ethernet physical layer (Ethernet PHY), analog-to-digital Converter (ADC), digital-to-analog Converter (DAC) and radio frequency (RF) chains. The components of the SDR are typically implemented by using a general-purpose DSP and FPGA. This allows the SDR to be programmed on the fly using many different communication protocols.
To meet the various communication requirements, customized SDR hardware with universal software radio peripheral (USRP) can be used to provide flexible platform design due to their lighter weight, low power consumption, wideband frequencies, and compact nature, such as N210 \cite{YeAir}, B210 \cite{GutierrezTime}, X310 \cite{BerghLTE, KhawajaUAV}, 2953R \cite{IzydorczykExperimental}, B200 mini and B205 mini \cite{CareemHiPER}.  In the past literature, USRP hardware has been extensively used in channel modeling \cite{YeAir, GutierrezTime, KhawajaUAV} and wireless communication platforms development \cite{BerghLTE, IzydorczykExperimental, CareemHiPER}, such as multi-carrier and multiple-input-multiple-output (MIMO) system in UAV-enabled communications. For example, a modular design allows the USRP N210 to operate up to 6 GHz \cite{YeAir}, while an expansion port allows multiple USRP series devices to be synchronized and used in a MIMO configuration \cite{IzydorczykExperimental}. The USRP application programming interface supports all USRP products and enables users to efficiently develop applications.
Each communication technology has its own advantages and disadvantages. For example, WiFi is a low cost solution, but has the main limitation of short range and vulnerability to interference. LoRa achieves low power and long range coverage, but with very low data rate. LTE is available almost worldwide with established infrastructure and technology, but a seamless 3D aerial coverage cannot be guaranteed in the sky at the moment. And SDR achieves versatile and low-power functionalities in a unified manner at the cost of high implementation difficulties. The users can select suitable communication technologies based on coverage area, communication distance, transmit power, data rate, and cost, etc. 
The comparison of different communication technologies is summarized in Table \ref{Communicationtechnology}.

\begin{table}[H] \footnotesize
\centering
\renewcommand{\arraystretch}{1.5}
\caption{Comparison of different communication technologies}
\label{Communicationtechnology}
\begin{tabular}{|p{2.4cm}|p{3.3 cm}|p{3.3cm}|p{1.5cm}|p{2.2 cm}|p{0.6 cm}|}
  \hline
  \textbf{Communication technology} & \textbf{Frequency band} & \textbf{Typical communication range} & \textbf{Transmit power} & \textbf{Customizability} & \textbf{Cost}\\
  \hline
  WiFi & 2.4 GHz, 5 GHz & LoS link & 29 dBm & No & Low\\
  \hline
  LoRa & 868 MHz, 915 MHz & 15 km for suburban area, 5 km for urban area & 5-20 dBm & No & Low\\
  \hline
  LTE & Selected cellular band & Virtually unlimited operation range due to the almost worldwide accessibility of cellular networks & 15-23 dBm & No & Low\\
  \hline
  SDR & 868 MHz, 915 MHz, 2.4 GHz, 5 GHz, all cellular band, and mmWave band & Hundreds of meters to thousands of meters & $ > $10 dBm & Yes & High\\
  \hline
\end{tabular}
\end{table}

\subsection{Communication protocol}
The widely used protocols for UAV communications include micro air vehicle link (Mavlink), UDP and TCP/IP, etc. Mavlink is a specialized communication protocol for UAV systems, which can be used to ensure the efficient communication between UAV and its GCS. Mavlink is an open-source protocol, specified for message structure and content, customized for UAVs in telemetry, command \& control information exchanging, as well as data transferring. It is now supported by all open-source and many close-source projects for UAVs. It supports sending way-points, control commands and telemetry data, switching flight modes, as well as adjusting parameters remotely. Another advantage of MavLink protocol is that
it supports different types of communication technologies, e.g., WiFi and LTE. An overview of Mavlink was presented in \cite{KoubaaMAVlink}, where all version 1, version 2 and their message formats were thoroughly presented, including main features of the Mavlink protocol, different tools and application program interfaces. They also surveyed the main contributions presented in the literature around Mavlink, including enhancement and extension, security, applications, integration with IoT, and swarm. An implementation of Mavlink protocol was given in \cite{ShiTestbedsMavlink}, where a robust control framework on a customized UAV platform was built using Mavlink. A USRP B200 mini, a Pixhawk2 flight controller, a embedded Linux system (Raspberry Pi3) to communicate with USRP B200 mini, and a 3DR Solo Quadcopter were used to realize the control information exchanging for UAV. The UAV autonomous scheme which allows for a customized control of mobility was demonstrated. The comparison of different communication protocols is summarized in Table \ref{Communicationprotocol}.

\begin{table}[H] \footnotesize
\centering
\renewcommand{\arraystretch}{1.5}
\caption{Comparison of different communication protocols}
\label{Communicationprotocol}
\begin{tabular}{|p{1.6cm}|p{1.6 cm}|p{2.6cm}|p{4 cm}|p{2.5 cm}|}
  \hline
  \textbf{Protocol} & \textbf{Reliability}  & \textbf{Packets} & \textbf{Messages arrive in order?} & \textbf{Application}\\
  \hline
  Mavlink & High &  Both streaming and datagrams & Select based on data type & CNPC and data transfer\\
  \hline
   TCP/IP & High  & Streaming& Yes & CNPC and data transfer\\
  \hline
   UDP & Low  & Datagrams& No & Data transfer\\
  \hline
  \multicolumn{5}{|l|}{CNPC: control and nonpayload communication}\\
  \hline
\end{tabular}
\end{table}

\section{Experiment for UAV A2G channel modeling}
The implementation of an advanced UAV communication system requires a comprehensive understanding of propagation channels between the UAV and ground nodes. Recently, there are some survey studies for A2G propagation channel experimental measurements and modeling \cite{Matolakair, KhawajaAsurveyof, MatolakUnmanned, KhuwajaAsurvey} of UAV communications. For instance, in \cite{Matolakair}, a comprehensive survey on wideband A2G channel was given to help the design of transmission schemes for UAV communications at L- and C-bands. Some basic definitions and importance of accurate channel modeling were also provided. In \cite{KhawajaAsurveyof}, the authors summarized the measurement campaigns for A2G channel modeling, including the type of channel sounding signal, operating frequency, transmit power, flight speed and altitude of UAV, link distance, elevation angle, and environment. The statistics of A2G channel were also provided. They described large-scale fading, small-scale fading, MIMO channel characteristics, simulation results, future research directions and challenges on A2G channel modeling. The authors in \cite{MatolakUnmanned} first described the basic A2G channel characteristics and limitations of the existing models, and then presented their A2G channel measurement campaign and provided example measurement and results on path loss and the Rician K-factor in a suburban or hilly environment. A survey of the channel characterization with measurement campaigns and statistical channel models was given in \cite{KhuwajaAsurvey}. The channel measurement campaigns were categorized as low altitude platform, low cost and low power solution, and widely deployed ground infrastructure. The empirical models were also reviewed. The UAV channel modeling methods were classified into deterministic, stochastic and geometric models. They further examined some challenging issues related to airframe shadowing, non-stationarity of channel, and diversity gain. In addition, several analytical channel models were comprehensively reviewed in prior work \cite{ZengAccessing}, which have been extensively adopted in the UAV communications research.
\begin{table}[tp] \footnotesize
\renewcommand{\arraystretch}{1.5}
\caption{Measurement Campaigns}
\label{Measurement Campaign}
\begin{tabular}{|p{0.6cm}|p{1.5cm}|p{2.6cm}|p{2.2cm}|p{1.6cm}|p{5cm}|}
  \hline
  \textbf{Ref.} & \textbf{Frequency} & \textbf{Equipment} & \textbf{Environment} & \textbf{Altitude} & \textbf{Channel Characteristics} \\
  \hline
  \cite{Matolak03} & 970 MHz, 5 GHz & 3B Viking aircraft & Over-Water & 580$\pm$13 m & PL, K-factor, RMS-DS, PDP, MPC, Intra- and inter-band correlation\\
  \hline
  \cite{Matolak04} & 968 MHz, 5.06 GHz & S-3B aircraft & Hilly Suburban & 602 m & PL, RMS-DS, PDP, Inter-band correlation \\
  \hline
  \cite{Matolak05} & 970 MHz, 5.06 GHz & S-3B aircraft & Near-Urban & 560 m & PL, K-factor, RMS-DS, Intra- and inter-band correlation \\
    \hline
    \cite{YeAir} & 2.585 GHz & Quadcopter, USRP N210 & Campus, & 0-300 m & MPC, RMS-DS, K-factor, CDF\\
    \hline
  \cite{NewhallWideband} & 2.05 GHz & - & Campus & 457.2-985.6 m &  PDP, RMS-DS, CDF, Diversity order\\
     \hline
   \cite{WillinkMeasurement} & 915 MHz &  Fixed\textendash wing, FPGA & Suburban & 200 m & RMS-DS, CIR, Spatial correlation, Spatial diversity  \\
     \hline
 \cite{GutierrezTime} & 5.8 GHz & DJI S-1000+ octocopter, USRP B210 & Residential, Mountainous Desert Terrains& Flat & PDP, RMS-DS \\
  \hline
  \cite{BerghLTE} & 800 MHz, 1800 MHz & Quadcopter, USRP X310 & Suburban & 0-120 m & RSRP, SINR \\ 
 \hline
  \cite{WangANovel} & 850 MHz & UAV, USRP & Campus & 15, 25 m & CIR, MPC \\ 
  \hline
  \cite{AmorimLTEradio} & 800MHz, 1800 MHz, 2600 MHz & Construction lift, Rohde-Schwarz radio scanner& Urban & 1.5-40 m & PL, CDF \\
  \hline
  \cite{LiuMeasurement} & 2.4 GHz & DJI Mavic 2 Zoom, Raspberry Pi 3b & Campus & 0-50 m & Throughput, Latency, RSSI, Pack loss \\ 
   \hline
 \cite{CaiAnEmpirical} & 2.585 GHz & Hexacopter, USRP N210 & Urban & 15-100 m & CIR, MRC, Delay, Doppler frequency\\
 \hline

 \cite{QiuLowAltitude} & 1.2 GHz, 4.2 GHz & Hexacopter & Suburban & 0-100 m &  PL, Height-dependent Rician K-factor\\
  \hline
  \cite{ShiMeasurement} & 900MHz, 1800 MHz, 5 GHz & DJI Matrice 100, USRP E312 &  LOS, NLOS area & 0-30 m & PL\\
 \hline
 \cite{AmorimRadio} & 800 MHz & Hexacopter, Rohde-Schwarz radio scanner & Urban & 15-120 m & Height-dependent PL exponent and shadowing variation \\
   \hline
   \cite{HouraniModeling} &  850 MHz & Quadcopter & Suburban & 15-120 m & Angle-dependent PL\\
    \hline
   \cite{RamírezWideband} & 1.2 GHz, 4.2 GHz & DJI Hexacopter & Semi-urban & 0-40 m & PDP, RMS-DS\\
  \hline
  \cite{AmorimPathloss} & 800 MHz & Cumulus One & Airport & 20-100 m & PL, SINR\\
   \hline
  \cite{KhawajaUWB} & 3.1-5.3 GHz & Tarot 650 quadcopter & Campus & 0-16 m & Large-scale and small-scale fading, MPC, CIR, foliage blockage, PDF\\
  \hline
 \cite{KhawajaUAV} & 28GHz, 60 GHz & DJI S-1000 octocopter, USRP X310 & Over Sea, Rural, Suburban, Urban& 2, 50, 100, 150 m & RSS, RMS-DS, MPC \\ 
 \hline
 \multicolumn{6}{|l|}{RSRP: reference signal received power, CDF: cumulative distribution function, CIR: channel impulse response}\\
\multicolumn{6}{|l|}{PDP: power delay profile, PL: path loss, RMS-DS: root mean square delay spread, RSS: received signal strength}\\
\multicolumn{6}{|l|}{PDF: probability distribution function, SINR: signal-to-interference-plus-noise ratio}\\
\hline
\end{tabular}
\end{table}
Different from the aforementioned surveys, in this subsection, we provide a survey on the prototype and experimental measurements on the UAV channel modeling to facilitate the design, evaluate, and optimize the coverage, reliability, and capacity performance of UAV communications. 

Efforts have been devoted in past literature to understand the A2G channel characteristics, which are mainly categorized into simulation-based method \cite{HouraniModelingair, HouraniOptimal, FengPathLoss, Alzenad3DPlacement} and measurement-based method. For the simulation-based methods, ray-tracing technique is generally used. The two-ray models, along with the simple free-space path loss model (which neglects any reflection or scattering) are simple analytical models. The free space path loss model is valid only when there is an unobstructed LOS path between the transmitter and the receiver and no objects in the first Fresnel zone. The two-ray model and  free-space path loss model are inaccurate (or at least incomplete) for settings where additional multipath components (MPCs) may be present.

In the following, we mainly focus on the measurement-based methods, which are summarized in Table \ref{Measurement Campaign}. Specifically, single-input-multiple-output (SIMO) channel measurement in the urban, suburban, hilly, and over sea scenarios on L- (970 MHz) and C-band (5 GHz) by using manned aircraft and transportable tower system on a trailer were provided in \cite{Matolak03, Matolak04, Matolak05}, respectively. The channel sounders were developed by Berkeley Varitronics Systems, Inc. Receivers with four antennas were installed at the bottom of the aircraft in a rectangular pattern and the transmitters at the ground station (GS). All antennas were vertically polarized. The time delay line model was employed to characterize the two-ray propagation with an additional intermittent mutipath component at altitude 580 m \cite{Matolak03}, 602 m\cite{Matolak04} and 560 m \cite{Matolak05}. Several channel characteristics were estimated by using data collected at channel sounder, e.g., path loss, delay spread, Doppler effects, small-scale fading characteristics, and correlations among the signals received on different antennas and in different bands.

Rotary-wing UAV and USRP-based measurements were performed in \cite{YeAir, GutierrezTime, BerghLTE, WangANovel, CaiAnEmpirical, ShiMeasurement, KhawajaUAV}. In \cite{YeAir}, a USRP N210 and a laptop were used to record the down-link signal with vertical polarization antenna. A feature selection algorithm was proposed to calculate channel parameters from CIRs. The results showed that the K-factor is the most height-sensitive parameter and can be modeled as a piece-wise function of height.
In \cite{GutierrezTime}, a DJI Spreading Wings S1000+ octocopter with payload consisted of an Intel NUC D34010WYK, a USRP B210, a iPhone 5S were used to build the transmitter. The iPhone 5S was used to record GPS information and time. The ground base station also consists of a USRP B210 mounted on top of an eight foot ladder. A MacBook Pro was used to record the transmitted data. A vertically polarized, dual band omni-directional vertical antenna with 3dBi gain were equipped at both transmitter and receiver. They studied the time and frequency dispersion characteristics of A2G channel at C-band (5.8 GHz) in outdoor residential and mountainous desert terrains by evaluating RMS-DS and Doppler spread.
In \cite{BerghLTE}, a sports airplane and a USRP X310 were used to digitize and record the LTE signals at 800 and 1800 MHz with altitude ranging from 150 m to 300 m. And GNU Radio, openLTE, and LTE Cell Scanner were used to analyze the recorded data. For lower altitudes between 0 and 120 m, a quadrotor UAV and a LTE phone running a LTE cell tracking application (G-MoN for Android) were used for measurement. The RSRP and SINR were used to characterize the A2G channel from UAV to ground LTE base station. The results showed that signals can be received from a large number of ground base stations as altitude increases, which leads to a decreased SINR at the UAV receiver.
In \cite{WangANovel}, a quasi-omnidirectional packaged discone antenna, a USRP device, a GPS-disciplined oscillator and a small computer were equipped at both the UAV and GS. And a pair of commercial WiFi routers were used in both UAV and GCS for remotely controlling. The channel measurement was conducted at 850 MHz in a suburban scenario at campus of Tongji University, Shanghai. The horizontal round-trip trajectory was planned with going altitude at 15 m and returning altitude at 25 m. The space-alternating generalized expectation-maximization algorithm was applied to estimate MPCs and analyze the concatenated PDPs. Further, to stuty environmental interactions on low altitude UAV A2G channel, simulation based on graph model is exploited to reconstruct the concatenated MPCs and PDPs. Similar measurement equipments, scenario and trajectory were adopted in \cite{CaiAnEmpirical} at frequency 2.585 GHz at the altitude of 15-100 m. Results showed that the K-factor is positively correlated with the altitude, the delay spreads and Doppler frequency spreads are negatively correlated with the altitude, and the path loss exponent decreases as the horizontal distance increases.
In \cite{KhawajaUAV}, a DJI S-1000 UAV and a USRP X310 were used for measurement of A2G channel at millimeter wave (mmWave) frequency bands (licensed band 28 GHz, unlicensed band 60 GHz) for four different environments: urban, suburban, rural and over sea. A block diagram using GNU radio for channel sounding was provided. They utilized the Remcom Wireless InSite ray tracing software and imitated the real time flight of UAV with a given trajectory to evaluate the channel behavior at mmWave channel. They analyzed RSS and RMS-DS of MPCs by changing altitude of UAV at different environments. The results showed that the RMS-DS was highly dependent on the altitude of the UAV as well as the density/height of the scatters around the UAV.
In addition, fixed-wing UAV was adopted in channel measurement \cite{Matolak03, Matolak04, Matolak05, WillinkMeasurement, AmorimPathloss}. FPGA was adopted in \cite{WillinkMeasurement}, Raspberry Pi 3b was used in \cite{LiuMeasurement}, and Rohde-Schwarz radio scanner was utilized in \cite{AmorimRadio}.

In summary, \cite{Matolak03, Matolak04, Matolak05, NewhallWideband, CaiAnEmpirical, RamírezWideband} focused on wideband channel measurement where \cite{CaiAnEmpirical} also studies narrowband channel measurement. From the perspective of diversity order, \cite{Matolak03, Matolak04, Matolak05} studied the SIMO scenarios, while \cite{NewhallWideband, WillinkMeasurement} investigated the MIMO scenarios and the others explored the SISO scenarios. Therefore, according to the results obtained from Table \ref{Measurement Campaign}, UAV channel characterization mainly depends on the propagation environment, operating frequency, channel sounding process, flying altitude, antenna orientation, placement position and flight dynamics. UAV A2G channels are usually more dispersive, incur larger terrestrial shadowing attenuations, and change more rapidly due to flight maneuvering. More comprehensive measurements are required for characterizing the A2G propagation with the environmental effects and the maneuvering of UAVs.

\section{Experiment for energy consumption model}
One of the key issues for UAV communications is the quite limited onboard energy of UAVs due to their SWAP constraint, which renders energy-efficient UAV communications extremely important. To this end, several previous works focus on the mathematical modeling of UAV energy consumption. Beyond energy consumption of signal processing, circuits, transmit and/or receive, and power amplification similar with terrestrial communications systems, UAV communication systems consume additional energy to remain aloft and flight. To optimize the energy efficiency of UAVs communications systems, the propulsion power consumption model is generally required. Studies of  propulsion power consumption based on real flight measurements can help mitigate the simulation bias and assess the energy efficiency in practical UAV communication environment.
In addition, the propulsion energy consumption of UAVs is typically much greater than communication-related energy consumption, which makes the analysis of energy efficiency much different from the conventional terrestrial communication systems.

Early works studying UAV energy consumption mainly focused on other applications rather than UAV communications, where empirical \cite{FrancoEnergy} or heuristic \cite{RichardsAircraft, MaMILP, Grtlipath, Meienergy} energy consumption models were usually developed. The authors in \cite{FrancoEnergy} derived an energy consumption model by performing a set of field experiments to understand the effects of flight speed, horizontal and vertical accelerations on energy consumption of a quadrotor UAV. Different experiments were performed. First, the UAV flew at the maximum acceleration and deceleration, and for every flight speed, the consumed power was obtained by multiplying the absorbed current by the voltage. Second, four different flight conditions were considered, e.g., horizontal flight, climbing, descending, and hovering, and the power consumption was modeled as a function of the speed. Based on the power consumption model obtained from experiments, they proposed an energy-aware path planning algorithm to minimize the energy consumption. However, no mathematical model on UAV energy consumption is derived, which makes the result difficult to be applied to other UAV models. In \cite{RichardsAircraft} and \cite{Grtlipath}, the energy consumption of UAV was modeled as a $L_1$-norm of the force, while it was modeled as a function of the square of the flight speed in \cite{Grtlipath}. However, no rigorous mathematical derivations were given for these heuristic models.

To this end, rigorous mathematical derivations were provided recently in our previous work \cite{Zengfixed} and \cite{Zengrotary} to obtain closed-form energy consumption models based on aerodynamics for fixed- and rotary-wing UAVs \cite{FilipponeFlight, JohnsonHelicopter}, respectively. For a fixed-wing UAV, the instantaneous power consumption was modeled as a function of flight speed ${\bf{v}}$ and acceleration ${\bf{a}}$, expressed as \cite{Zengfixed}
\begin{align}\label{eq:UAVPOWER}
{P_C} = \left| {{a_1}{{\left\| {\bf{v}} \right\|}^3} + \frac{{{a_2}}}{{\left\| {\bf{v}} \right\|}}\left(1 + \frac{{{{\left\| {\bf{a}} \right\|}^2} - {{({{\bf{a}}^T}{\bf{v}})}^2}/{{\left\| {\bf{v}} \right\|}^2}}}{{{g^2}}}\right) + m{{\bf{a}}^T}{\bf{v}}} \right|,
\end{align}
where $a_1$ and $a_2$ are constants that are independent of the flight status but only related to aerodynamics and aircraft design, such as air density, zero-lift drag coefficient, wing area, wing span efficiency, aspect ratio of the wing, and total weight of aircraft, $m$ denotes the mass of UAV, $g$ is the gravitational acceleration in $m/s^2$, and $\left|\bullet\right|$, ${(\bullet)^T}$, and $\left\|\bullet\right\|$ denote magnitude operator, transpose operator, $L_2$-norm of a vector, respectively. 

For a rotary-wing UAV, the derivation of energy consumption model is much more complicated than the fixed-wing UAV. Our previous work in \cite{Zengrotary} only derived the power consumption model following straint-and-level flight trajectory, which can be expressed as
\begin{equation}\label{eq:UAVLINEPOWER2}
P_{C} = {P_0}\left( {1 + \frac{{3{v^2}}}{{U_{{\rm{tip}}}^2}}} \right) + {P_i}{\left( {\sqrt {1 + \frac{{{v^4}}}{{4v_0^4}}}  - \frac{{{v^2}}}{{2v_0^2}}} \right)^{1/2}} + \frac{1}{2}{d_0}\rho sA{v^3},
\end{equation}
where ${P_0}$ and ${P_i}$ represent the blade profile power and induced power in hovering status, ${d_0}$, $\rho $, $s$, $A$, ${U_{{\rm{tip}}}}$, and ${v_0}$ are constants corresponding to aerodynamics and aircraft design, see Table I in \cite{Zengrotary} for reference. 
With the help of the derived models provided in \cite{Zengfixed} and \cite{Zengrotary}, a large mount of theoretical studies have been performed to optimize the energy efficiency of UAV communications, such as UAV base station, UAV relay, UAV data collection/dissemination, UAV multicast, UAV physical layer security, UAV data offloading.

\begin{figure}[H]
\centering
\includegraphics[width=12cm]{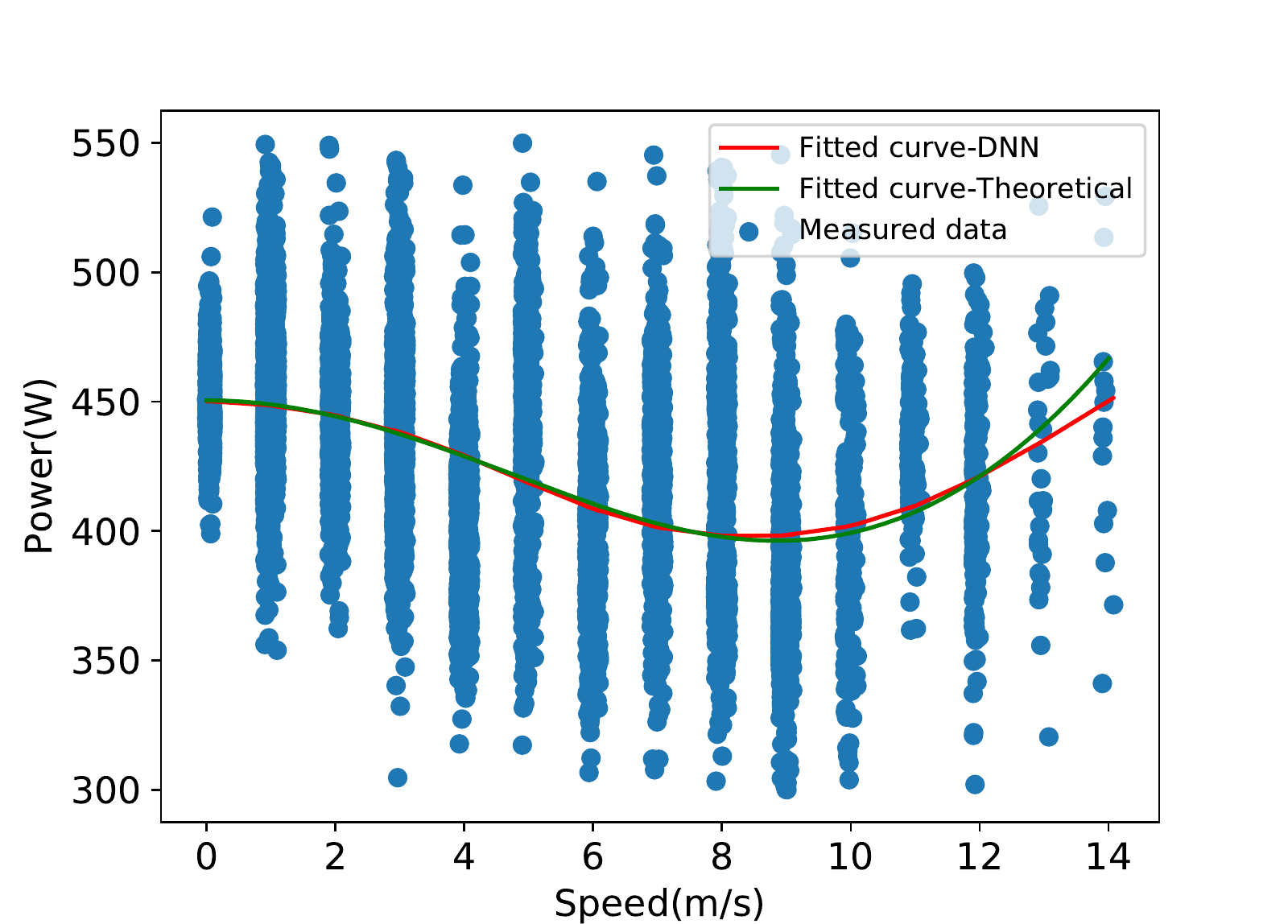}
\caption{Experimental verification for rotary-wing UAV energy consumption model \cite{{GaoEnergyModel}}.}
\label{energy}
\end{figure}

Recently, we have performed flight experiment to validate the energy consumption model of rotary-wing UAV \cite{{GaoEnergyModel}}. In the flight experiment, the instantaneous current and voltage of on-board battery, as well as the flying status of the UAV (e.g., location, speed, and acceleration) were recorded. Based on these collected data, we applied the model-based curve fitting method to obtain the modelling parameters in (\ref{eq:UAVLINEPOWER2}), as well as a model-free deep neural network (DNN) training to exclude the potential bias caused by the theoretical model. As illustrated in Figure \ref{energy}, the obtaied curve from both methods match quite well with each other, which validate the energy consumption model given in (\ref{eq:UAVLINEPOWER2}).

However, the power consumption models derived in \cite{Zengfixed} and \cite{Zengrotary} are only applicable to UAV flight in 2-D plane. For arbitrary 3-D UAV trajectory with climbing or descending over time, a heuristic closed-form approximation was proposed in \cite{ZengAccessing} and \cite{{GaoEnergyModel}}, but no rigorous mathematical derivations been performed. In addition, some ideal conditions and approximations have been assumed and adopted, such as zero wind speed. The energy consumption model by considering the effect of wind speed remains challenge.
In addition, the theoretical models obtained in \cite{Zengfixed} and \cite{Zengrotary} have not been completely validated by flight experiments and measurement.
\begin{figure}[H]
\centering
\includegraphics[width=12cm]{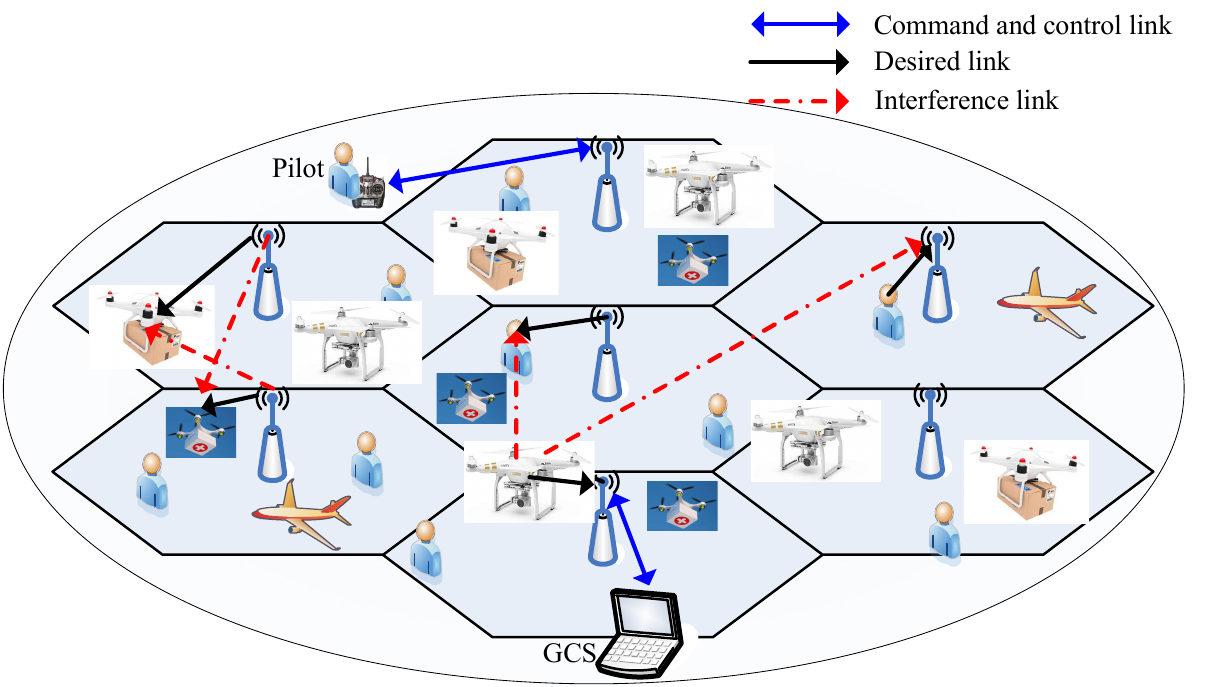}
\caption{Illustration of cellular-connected UAV.}
\label{aerialuser}
\end{figure}
\section{Cellular-connected UAV}
Cellular-connected UAVs have a great potential for search and rescue, inspection, entertainment and media, as well as traffic monitoring, etc., where UAVs are integrated as new aerial users that access the cellular network from the sky \cite{ZengAccessing, VinogradovTutorial, MozaffariATutorial, BerghLTE, 3GPP}. Thanks to the almost worldwide availability of terrestrial cellular networks, it is possible for the UAV operators or pilots to control the UAV remotely with beyond LoS links. In addition, cellular-connected UAV provides a cost-effective solution since it may utilize the existing terrestrial cellular base stations without the need of building new dedicated infrastructures exclusively for supporting UAV communications. To integrate UAVs as aerial cellular users, it needs to provide reliable and low-latency communication links for exchanging command \& control messages between ground pilots and UAVs with the help of cellular networks, while supporting various payload communication requirements for specific applications. However, there are still several critical challenges to be addressed before current cellular networks can be used to provide full support for UAV users. In particular, as current cellular networks were mainly designed to serve ground users, whose channel characteristics, mobility, operation altitude are quite different from that of UAV users, a seamless 3D coverage in the sky cannot be guaranteed by existing cellular networks. Besides, extensive experiments and simulations have revealed that the severe aerial-ground interference is a another critical issue for cellular-connected UAVs\cite{3GPP, Muruganathan3GPP, BerghLTE, QualcommLTE, IvancicFlying, JwaNetwork, YangLTEAdvanced, KovacsInterference, BatistatosLTE}, as illustrated in Figure \ref{aerialuser}.

In the following, we first investigate the feasibility of cellular-connected UAV by reviewing the existing measurement campaigns. And then, we review the experiments for addressing the main challenges of interference mitigation for cellular-connected UAVs.

\subsection{Feasibility study of cellular-connected UAV}
In general, the antennas of cellular base stations are downtilted towards the ground to cover the associated ground users and mitigate the intercell interference \cite{ZengAccessing}. Due to the high flying altitude and thus the increased LoS probability, UAVs are normally capable of receiving signals from several base stations via their side-lobes.
Recently, several experiments were performed to validate the feasibility of providing wireless connectivity for UAVs by utilizing the existing LTE networks \cite{QualcommLTE, BerghLTE, SaePublicLTE, BatistatosLTE, IvancicFlying, JwaNetwork, YangLTEAdvanced, KovacsInterference}.

For example, both measurements and simulation results were provided in \cite{BerghLTE} to study the impact of flying altitude, horizontal distance from the ground base station, and UAV density on the performance of cellular-connected UAV. It was found that as the UAV altitude increases, the number of detectable base stations at the UAV increases, and the received power at the cellular-connected UAV gets stronger. However, UAV's received SINR degrades, mainly due to the increased interference. They concluded that interference is a major limiting factor for cellular-connected UAV. Such observations have also been corroborated by other field measurement campaigns \cite{QualcommLTE, IvancicFlying, JwaNetwork, YangLTEAdvanced, KovacsInterference}. In the field trials of Qualcomm, hundreds of flights were performed on three LTE bands (PCS, AWS, and 700 MHz) at altitude below 120 m to validate the safe operation of the UAV, the completeness and correctness of the logged data, as well as to collect the data sets for final analysis\cite{QualcommLTE}. They claimed that commercial LTE networks should be able to support downlink communication of initial cellular-connected UAV without major change. For uplink communications, aerial users causes more severe uplink interference than ground users since free space propagation intensifies the interference received at neighbor base stations. With their measurement configurations, aerial UAV produced approximately 3 times the interference than a ground user in 700 MHz. However, this effect should not be a problem for initial deployment of cellular-connected UAV with some of them supporting high speed uplink transmission. In addition, measurements in \cite{JwaNetwork} showed that it is feasible for command \& control messages exchanging with flying altitude up to 150 m but the requirement of high speed transmission can not be met at high flying altitudes.

In \cite{SaePublicLTE}, the measurements for public cellular network in a rural area to cover coexisting aerial users and ground users were performed, which aim to study the impact of aerial users interference on the ground users. Four key performance metrics were presented, i.e., physical resource block (PRB) usage, modulation and coding scheme (MCS), throughput and transmission power. They observed that the aerial users reduce the MCS class of the ground users, with roughly the same uplink throughput. In addition, from purely uplink throughput point-of-view, integrating aerial users into cellular network does not have critical impact on the performance of the normal ground users, but at the cost of increased PRB usage and transmit power. Moreover, the users closer to the cell edge suffered severer interference from aerial users than those closer to the base station.

In summary, although the antennas of cellular base stations are downtilted towards the ground, extensive measurement campaigns have demonstrated the feasibility to provide connectivity for UAV users for initial deployment at low altitude (say below 122 m) by using the existing ground cellular networks \cite{QualcommLTE, YangLTEAdvanced, SaePublicLTE}. However, this usually comes at the cost of increased interference in both uplink and downlink communication, PRB usage and transmit power. Therefore, mitigating the interference induced by cellular-connected UAV is the most crucial issue to improve the performance of cellular-connected UAV. However, providing seamless connectivity for moderate and high altitude using existing cellular networks is still a big challenge. Moreover, supporting the communication for the coexisting densely deployed UAV and ground users needs to be further investigated.

\subsection{Interference mitigation}\label{Interference}
Conventional techniques for uplink and downlink interference mitigation uses orthogonal channel access, like FDMA and TDMA , which, however, usually results in poor spectrum utilization. To this end, many advanced techniques have been developed \cite{3GPP, Muruganathan3GPP}. For example, power control, directional antenna and MIMO beamforming can be used to mitigate the uplink interference caused by the communication link from UAVs to neighboring base stations. For the mitigation of downlink interference from cochannel base stations to UAVs, directional antenna and MIMO beamforming can be similarly applied. 
Recently, various techniques have been explored in measurement campaigns to mitigate the interference induced by cellular-connected UAV, e.g., power control \cite{NguyenHowto}, directional antenna \cite{3GPP, Lidirectional}, multi-antennas \cite{IzydorczykExperimental, NguyenHowto, VinogradovTutorial, HeimannmmWave}, and joint cooperative multi-point transmission \cite{NguyenHowto, KovacsInterference}, etc.

Terminal-based and network-based interference mitigation techniques were analyzed in \cite{NguyenHowto}. For terminal-based solutions, it is observed that interference cancellation and antenna beam selection can improve the performance of both cellular-connected UAVs and ground users with up to 30\% throughput gain, and increase reliability of UAV connectivity to above 99\%. For network-based solutions, the uplink power control and downlink joint cooperative multi-point transmission have been studied. The results revealed that the uplink power control technique can improve the average uplink throughput of ground users. For heavily-loaded scenarios, the downlink joint cooperative multi-point transmission can provide the required downlink performance at the cost of 10\% performance degradation of ground users in the associated and affected cells.

Directional antenna is another promising technique for interference mitigation for cellular-connected UAVs. For directional antenna with fixed radiation pattern, the antenna gain is a deterministic function of the elevation and azimuth angles, and the main lobe gain is usually much larger than that of the side lobes. The performance of cellular-connected UAVs can be greatly improved by aligning the antenna main lobe with the direction of the associated base station, which can be achieved either mechanically or electrically (multi-antennas beamforming). A mechanically automatic alignment system of directional antennas was developed in \cite{Lidirectional}, where the communication quality indicator and RSSI were used for the objective function of antenna alignment. This antenna alignment system can be directly applied in interference mitigation for cellular-connected UAVs. 

To control the antenna radiation pattern electrically, multi-antennas beamforming is a natural solution. Beamforming is an effective technique that can adjust the antenna radiation pattern dynamically towards the desired directions. This significantly improves the performance of interference mitigation. In \cite{IzydorczykExperimental}, a cellular-connected UAV platform for beamforming experiment was built and used in a measurement campaign. The results showed that multi-antenna beamforming can extend the signal coverage, reduce interference and reduce handover occurrences. To further enhance the throughput of cellular-connected UAV, future cellular techniques like mmWave and massive-MIMO need to be studied in depth to reduce interference and enhance the performance.

\section{UAV-enabled communication platform}
\begin{figure}[tp]
\centering
\includegraphics[width=12cm]{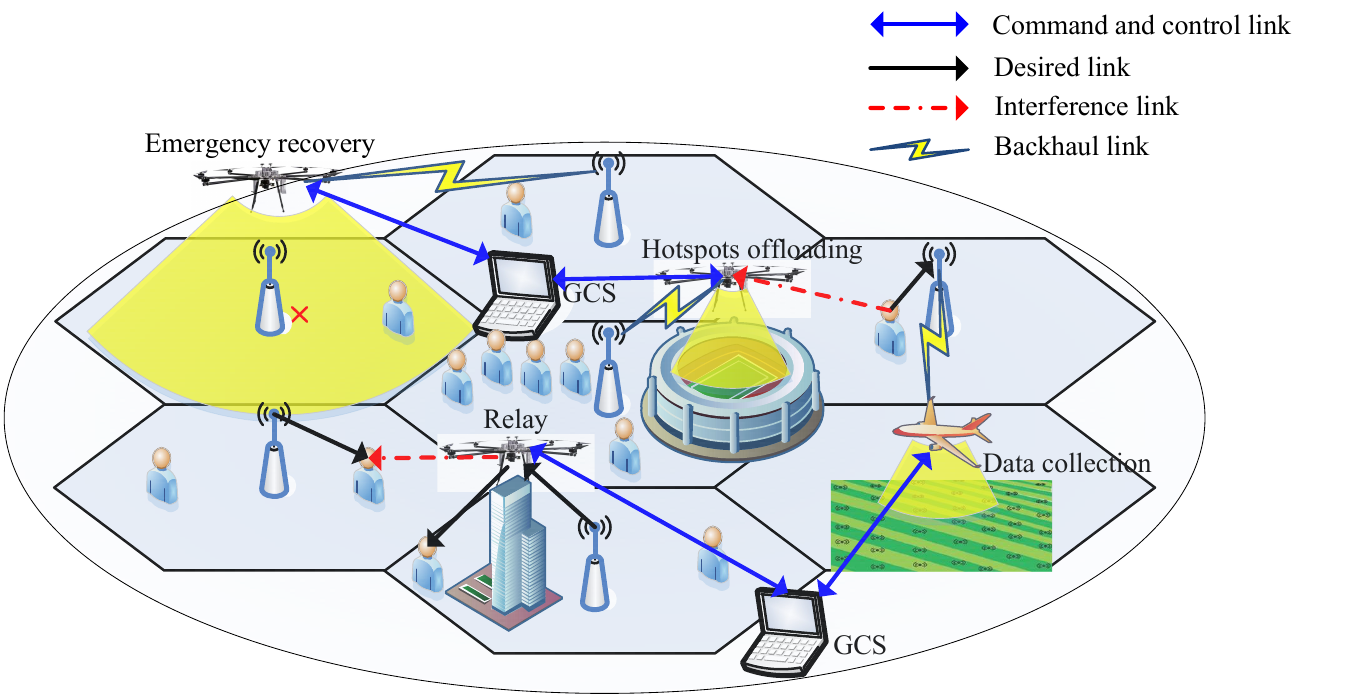}
\caption{Illustration of UAV-enabled communication platform.}
\label{platform}
\end{figure}

In this section, we focus on the other paradigm, namely UAV-enabled communication platform, which aims to provide aerial wireless access for terrestrial users from the sky. Due to the highly controllable mobility, adjustable altitude, fast and on-demand deployment capability, as well as the unique characteristics of A2G channels, UAV-enabled communication platform provides a promising complementation of existing terrestrial communication systems, which has attracted significant attention from both industry and academia. In fact, by utilizing these unique characteristics, UAV communications can significantly enhance the performance of existing terrestrial communication systems, including coverage area, throughput, delay, and overall quality-of-service. As illustrated in Figure \ref{platform}, UAV-enabled communication platforms have three typical use cases: UAV-enabled base station, UAV-enabled relaying and UAV-enabled data cellection/dissemination \cite{{ZengWireless}}. Specifically, UAV-enabled base station aims to provide wireless coverage for the targeting geographical areas with limited or no cellular infrastructure. Typical scenarios include temporary data offloading in hot spots, and fast post-disaster communication service recovery. UAV-enabled relaying can be used to extend coverage area of cellular base station, establish communication link between distant users without direct link. And UAV-enabled data cellection/dissemination can be employed as aerial access points to collect/desseminate data from/to ground nodes in WSN and IoT communications. It is worth mentioning that UAV-enabled communications platform also faces severe interference. The techniques discussed in Section \ref{Interference} can be also used for interference mitigation in UAV-enabled communications platform.

\subsection{Experiment for UAV-enabled aerial base station}
Recently, UAV-enabled aerial base stations have attracted significant attention from industry. For example, Verizon has presented an airborne LTE operations project \cite{{VerizonALO}}, with the potential services including: inspection of pipelines and high-voltage power lines without endangering people, aerial imaging to increase agricultural yields of farmland, and unmanned broad views of wildfires or storm damage by first responders. Several controlled trials in New Jersy using large 300-pound UAV were performed to carry LTE cellular equipment into the sky. These trials were aimed to create an in-flight cellular network to replace traditional network \cite{{VerizonALOemergencies}}. In addition, the project of cell on wings (C OW) was performed by AT\&T, where Flying COW was built to beam LTE coverage from the sky to ground users during disasters or big events \cite{ATTCellOnWings}. A successful test flight of Flying COW transmitting and receiving high speed data was performed above a field outside Atlanta. The tested tethered UAV carried a small cell and antennas. The tether between the UAV and the ground provided a data link via fiber and supplied power to the Flying COW, which allowed for unlimited endurance. For the ongoing 5G communication system, Qualcomm and AT\&T are planning to deploy UAV-enabled communication platform for large-scale communications in 5G wireless networks \cite{AirborneLTE}. Huawei together with China Mobile performed a test of high altitude UAV 5G base station for emergencies \cite{HUAWEI5G}, in which the tethered UAV, 4G/5G dual mode BOOKRRU and customized omnidirectional antennas were used to achieve a maximum coverage of 6.5 km at altitude 200 m.

From the perspective of academia research, a few measurements were performed \cite{BerghLTE, NunnsAutonomous, ChandrasekharanDesigning}.
In \cite{NunnsAutonomous}, the authors implemented a UAV access point which can adaptively adjust their position depending on the movement of the ground users. They continuously kept track of the RSS from the ground users for the estimated distance between UAV and ground users. They demonstrated that the UAV was able to localize ground users and autonomously adjust its position to move closer to them. However, the wireless backhaul was not considered. 
In \cite{BerghLTE}, the authors quantified the impact of LTE-enabled UAVs on an existing ground LTE network for dual cases by using a combination of simulations and measurements, i.e., cellular-connected UAV and UAV-enabled aerial base station.
In particular, UAV-enabled aerial base station employed an omnidirectional antenna and operated with reduced transmit power, which ensured that the UAV transmits the signal in all directions, and the reduced transmit power limit the interference on neighboring cells. The results showed that the UAVs have a high impact in a significant portion of the coverage area and interference to the macrocells even with a reduced transmit power. The area covered by aerial base stations increases as the number of UAVs increases. 
The ABSOLUTE project presented in \cite{ChandrasekharanDesigning} employed a Helikite to implement an aerial base station, the satellite-based backhaul was adopted. The Helikite was a special combination of kite and balloon which uses both helium and wind to produce lift.

Although experiments have been performed by both industry and academia, UAV-enabled aerial base station is still in its initial stage. Most of the aforementioned works use satellite-based backhaul, which is usually costly and incurs high latency. Other works mainly utilize the tethered UAV to provide power and backhaul via fibre, which significantly limit the unique feature of UAV's 3D mobility. In addition, two categories of UAV were mainly employed, i.e., large UAVs operate at high altitude and quasi-stationary tethered UAVs operate at low altitude. Therefore, for low-altitude aerial base station in real-time applications, high-capacity, cost-effective and low latency wireless backhauling needs to be established between aerial base station and the core network on the ground, e.g., mmWave and relay.

\subsection{UAV-enabled aerial relay}
UAV-enabled aerial relays can be used to establish or enhance the connectivity between ground users (or ground users and base station) that cannot directly communicate with each other due to obstacles or long distance. They have been widely used for cellular coverage extension, wireless backhaul, and emergency response, etc. A few experiments have been conducted to validate the superiorities of UAV-enabled aerial relay \cite{GangulaFlyingRebots, OnoMeasurement, GuoPerformance}.

For example, the authors in \cite{GangulaFlyingRebots} used a COTS Quad-Rotor carbon body frame, DJI propulsion system and open source PIXHAWK 2 flight controller to build a customized integrated UAV relay  between the ground user and a fixed base station based on OpenAirInterface. They presented an autonomous placement algorithm that updates the 3D position of UAV relay in realtime to maximize the downlink throughput according to the user location, 3D map of the environment and propagation channel. The results showed that the optimal deployment location can often achieve LoS links between the user and the fixed base station. However, the radiation and the polarization of the antenna, and the impact of UAV altitude as well as its antenna orientation have not been taken into account.

In \cite{OnoMeasurement}, the authors conducted an experimental measurement for fixed-wing UAV-enabled aerial relay and evaluated the performance of TCP and UDP protocols for both non-urban area and urban area. The back and forth trajectory to ferry data was adopted. The results showed that the UDP protocol outperforms the TCP protocol in terms of throughput. However, the impact of antenna radiation pattern, UAV altitude and resource allocation have not been considered.

In \cite{GuoPerformance}, the authors presented experimental field trials of UAV-enabled aerial relay test-bed in both rural and urban environments. They employed a Parrot Drone Mk2 and the decode-and-forward strategy. The channel between base station and UAV was a 3G in-band channel, and that between UAV and user was the out-of-band WiFi channel. The altitude of UAV was below 15 m. The results showed that throughput improvements can be achieved for users in poor coverage zones.

In summary, there are limited works focusing on the experiment for UAV-enabled aerial relay. To facilitate the practical application of UAV-enabled aerial relay, more efforts are still needed, such as trajectory optimization to maximize the throughput with a given energy budget or minimize the energy consumption with given amount of data, 3D deployment location optimization to maximize the throughput, multiple UAVs-enabled multi-hop relay, and multi-antennas beamforming for UAV relay, in various operation environments.

\subsection{UAV-enabled aerial data collection/dissemination}
With their 3D mobility and unique channel characteristics, UAV-enabled data collection/dissemination is expected to play an important role for WSN and IoT communications  \cite{MoheddineUAV, LiangForest, PopescuCollaborative, CaoDesign}, such as smart cities infrastructure management, healthcare, precision agriculture/forest, and energy management.

In \cite{MoheddineUAV}, the LoRa-based gateway was deployed on a UAV flying over IoT nodes to collect temperature and humidity data and transmit them to a LoRa server. A Rak2245 Pi-Hat module and a Raspberry Pi were used to build the gateway for providing LoRa connectivity which usually achieves the advantages of long coverage and low power consumption. Each IoT node consists of a sensor and an antenna. Both Ethernet/WiFi and cellular network were tested. However, only two ground nodes were used and the impact of altitude was not considered.

The precision agriculture/forest scenarios were studied in \cite{PopescuCollaborative, LiangForest}. The authors in \cite{PopescuCollaborative} utilized fixed-wing UAV to collect data of precision agriculture and relay them to a data center efficiently through trajectory optimization and in situ data processing. In \cite{LiangForest}, the UAV collected the forest data to help the land owners for mapping the forest without professional equipment and knowledge.

The authors in \cite{CaoDesign} constructed a UAV platform to collect the data of WSN. Tests both on communication capacity and data collection were carried out. They adopted a fixed-wing UAV with materials of foamed polypropylene. A CC2530 wireless module operating at 2.4 GHz and a 3 dBi omni-directional sucker antenna were used. In particular, both tension, rotational speed
and wind field were tested at UAV. The results showed that the UAV-enabled data collection can maintain a high quality communication link with lower packet loss rate and better RSSI at data rate 250 kbps, when the transmit power was greater than 1 dBm and the distance was less than 100 m.

In summary, the experiment for UAV-enabled aerial data collection/dissemination still lies in the early stage. To facilitate the practical application of UAV-enabled aerial data collection/dissemination, more efforts are needed to enhance the performance, such as trajectory optimization to minimize the completion time or energy consumption of both UAV and ground nodes, backscatter communication of WSN or IoT for energy and data transfer simultaneously, combination of device-to-device communication and UAV-enabled aerial data collection/dissemination.

\section{Extensions and future work}

\subsection{Channel modeling}
Experimental measurements on the UAV channel modeling are beneficial for the design, evaluation, and optimization of the coverage, reliability, and capacity performance of UAV communications. Although extensive measurement campaigns have been considered, there are several open issues need to be further investigated. First, most measurement efforts focused on urban, suburban and open fields environments with mostly clear LoS conditions, whereas, measurement efforts are still missing for scenarios with different building shapes, densities, streets, trees, lake water, as well as weather conditions. Second, most works studied channel modeling for sub-6G Hz band, little efforts were devoted to channel modeling on mmWave band. Compare with sub-6G Hz band, mmWave band suffers from more severe attenuation, and easier to be bloked. To understand propagation characteristics of mmWave A2G channel (e.g., multi-frequency comparison, blockage effects, weather effects, atmospheric attenuation, environment effects and massive MIMO properties), more efforts are required. Furthermore, the impact of airframe shadowing, non-stationarity of channel, antenna orientation, and diversity gain of MIMO channels should be well exploited via experimental measurements. Finally, almost all the measurement campaigns assumed the static ground receivers, the impact of ground receivers' mobility has not considered, which leads to a more complicated relative motion.

\subsection{5G cellular-connected UAV}
There is a growing demand for high throughput, ultra-reliable low-latency communications and massive connectivity for 5G cellular-connected UAV communication, and this will be challenging for current cellular networks. Several advanced wireless techniques have been involved in 5G communication network, e.g., massive MIMO and mmWave. For measurement experiment of 5G cellular-connected UAV, several issues should be taken into account, such as 3D mobility, 3D beamforming, interference mitigation, handover management, energy efficiency, potential large-scale deployment and high throughput applications. Due to the much smaller wavelength of mmWave, it is possible to integrate more antennas with compact size on UAV to exploit the gain of multi-antenna 3D beamforming. 3D beamforming is an effective multi-antenna technique that can dynamically adjust the antenna radiation pattern based on the UAV's real-time location thus significantly improving the throughput and mitigating the interference. However, no measurement campaigns of 3D beamforming have been performed. In addition, the maximum density that the 5G cellular network can support should be tested before large-scale deployment of cellular-connected UAVs. Furthermore, advanced interference mitigation techniques are required. The 3D mobility of UAV and denser deployment of 5G cellular base stations may incur more frequent handovers. In order to maintain reliable connectivity of UAV, an in-depth understanding of such handover behavior is of significant importance.
Besides, caching is another promising technique for 5G-and-beyond cellular networks, which can also be applied to improve the users' quality-of-experience (QoE) and reduce the transmission delay for cellular-connected UAV applications \cite{Cachingchen, ChenLiquid}. The experiments of caching-enabled cellular-connected UAVs are an interesting direction worth further investigation in the future.

\subsection{Trajectory optimization for UAV communications}
Due to the controllable high-mobility of UAV, trajectory optimization is one of the most important topics for UAV communications. Trajectory optimization offers an additional degree of freedom for communication performance enhancement by dynamically adjusting the UAVs' 3D locations. In practice, the trajectory optimization for UAV communications are constrained by UAV flight constraints (e.g., minimum/maximum flying altitude, initial/final locations, maximum/minimum UAV speed, maximum acceleration, obstacle avoidance, collision advoidance, and no-fly zone), and the communication-related constraints, (e.g., transmit power, time, and frequency resource limitations). Furthermore, the trajectory optimization problem is usually nonconvex with respect to communication and trajectory variables, and involves infinitely many continuous-time variables, thus making it difficult to be directly solved. In the literature, there have been extensive theoretic studies that applied different approaches to solve this problem. For instance, a widely adopted method is to first use the classic travelling salesman problem and pickup-and-delivery algorithm to obtain the initial UAV trajectory, and then adopt the time/path discretization together with block coordinate descent and successive convex approximation techniques to optimize the trajectories (see, e.g., \cite{ZengAccessing}and the references therein). To the authors' best knowledge, the trajectory optimization for UAV communications still lies in stage of theoretic studies, and the experiment campaigns to practically validate the optimized trajectory designs are still lacking. Besides, to facilitate practical application of trajectory optimization for UAV communications, the online optimization algorithms with low computation complexity are required to implement on an embedded platform.

\subsection{Energy-efficient UAV communications}
Due to the SWAP constraint of UAV, energy-efficient UAV communications is of utmost importance. Rigorous mathematical derivations of energy consumption model for both fixed- and rotary-wing UAVs have been given in our previous works \cite{Zengfixed, Zengrotary}. However, we have not considered the impact of the speed and direction of wind, the propulsive efficiency (the fraction of the engine power into useful power), and weather conditions, which may affect the total energy consumption of UAVs. Therefore, extensive measurement campaigns should be performed to obtain more practical energy consumption model. Based on the obtained energy consumption model, it will help to more effectively design energy-efficient UAV communication systems. In addition, our previous works mainly focused on 2D UAV flight with constant altitude, the impact of 3D mobility on energy consumption model has not been taken into account \cite{Zengfixed, Zengrotary}, and the impact of acceleration of rotary-wing UAV also has not been considered \cite{Zengrotary}. Furthermore, energy-efficient trajectory optimization still lies in theoretical analysis stage, no measurement campaigns have been conducted to validate the effectiveness of the proposed algorithm. And most of the current work on trajectory optimization adopted free space path-loss model, the impact of more practical channel model on energy efficient UAV communication has not been validated via measurement experiments. Moreover, for massive-MIMO equipped at UAV, the communication-related energy consumption can not be ignored. Furthermore, recent antenna technique such as large-scale passive intelligent reflecting surfaces can be adopted in UAV communication to help gain higher diversity order and enhance the energy efficiency \cite{LuEnabling}.

\subsection{Machine learning-based UAV communication}
Machine learning is a promising technique which can significantly improve the performance of UAV communication by autonomously learning from operation environment and their past behaviors {\color{blue}\cite{ChenArtificial}}.
Machine learning can be potentially utilized to design and optimize UAV communication systems. For example, UAVs can rapidly and autonomously adapt to dynamic environments, and optimize their 3D locations, trajectories to provide coverage for ground users. In addition, by leveraging machine learning algorithms, UAV-enabled aerial platforms can predict the ground users' mobility behaviors and their distributions, which can be used to enable optimal deployment of aerial platform. For the case of cellular-connected UAV, the ground base stations can predict the moving directions and flying speed of UAV, and adjust their beam to align with UAV. Furthermore, machine learning can also be used to learn the radio environments, which can be used to build a 3D A2G channel model. Such 3D A2G channel model can be subsequently used to design, evaluate, and optimize the UAV communication. Another future directions of machine learning are multi-UAVs collision avoidance and collaboration. The prototype and experiment of machine learning-based UAV communication for aforementioned directions deserve further investigations.

\section{Conclusions}
In this paper, we have provided a comprehensive survey on the prototype and experiment for UAV communications. We first discussed the general architecture for UAV experiment, including aircraft selection, communication technologies and communication protocol, and then present experimental verification for air-to-ground channel models. Next, to facilitate the design of energy-efficient UAV communication, we studied the UAV energy consumption model of both fixed- and rotary-wing UAVs. In addition, we explored the state-of-the-art experiments for two main paradigms, i.e., cellular-connected UAVs and UAV-enabled communication platforms. We also highlighted the extensions and some promising future directions in experiments and prototype of UAV communications that deserve further investigations.

%
%






\end{document}